\documentclass{elsart1p}
\usepackage{graphicx}
\usepackage{amssymb}
\begin{document}
\begin{frontmatter}
% use the corauthref command within \author for corresponding author
% footnotes;
\title{Symmetry, Confinement and the phase diagram of QCD}
\author{Adriano Di Giacomo}
\address{Pisa University and INFN Sezione di Pisa}
\begin{abstract}
A general discussion is presented of the possible symmetries responsible for confinement of color and of their evidence in lattice simulations. The consequences on the phase diagram of  $QCD$ are also analyzed.
\end{abstract}
\end{frontmatter}
%
% main text
\section{Why symmetry?}
\label{1}
 No free quark has ever been observed in Nature:  the abundance of quarks $n_q$ compared to the abundance of protons $n_p$ has an experimental upper bound ${n_q\over n_p}\le 10^{-27}$ to be compared to the value $10^{-12}$ expected in the Standard Cosmological Model in absence of confinement. The cross section for inclusive production of quarks in hadron collisions, $\sigma_q$ is also $10^{-15}$ times smaller than the perturbative expectation. The natural explanation of these facts is that  confinement is an absolute property, in the sense that $n_q$ and $\sigma_q$ are strictly zero due to some symmetry.
 As a consequence the deconfining transition is a change of symmetry, i.e. an order-disorder transition and can not be a cross-over.
 A similar situation exists in ordinary superconductivity: the resistivity in the superconducting phase  has an  exceedingly small experimental upper limit compared to the resistivity in the normal phase. The natural explanation is that the resistivity in the superconducting phase is strictly zero. A change of symmetry occurs at the transition from a Higgs broken $U(1)$ symmetry (superconductor) in which Cooper pairs condense in the vacuum, to a normal phase in which the $U(1)$ symmetry is exact.
 \section{What symmetry?}
 \label{2}
  Color symmetry is exact :  it can not distinguish confined from deconfined. Center symmetry only exists in absence of dynamical quarks. Chiral symmetry only exists at zero $m_q$. Moreover in some cases like $QCD$ with $N_f=2$ adjoint fermions chiral symmetry restoration occurs at a different temperature than deconfinement \cite{k}\cite{cos}, indicating that the relevant degrees of freedom  at the deconfining transition are not the chiral ones.
  The only way to get an extra symmetry is via duality \cite{KW}\cite{W}, i.e. by looking at excitations with topologically non trivial boundary conditions.
  In $(2+1)dim$ the homotopy is $\Pi_1$ and the topologically non trivial excitations are vortices, in $(3+1)dim$ the homotopy is $\Pi_2$ and the excitations are monopoles \cite{'tH}\cite{Pol}.
   For a generic gauge group $G$ of rank $r$, $r$ abelian field strength tensors ('t Hooft tensors)  $F^a_{\mu \nu}$, $(a=1,..r)$ can be defined\cite{dfp} and in terms of them $r$ magnetic currents 
 $j^a _{\nu} \equiv \partial_{\mu}F^{a*}_{\mu \nu}$ .
 Non zero value of the currents $j^a_{\nu}$ is a violation of Bianchi identities, due to the  presence of magnetic charges. The currents $j^a_{\nu}$ are conserved due to the antisymmetry of the dual tensor 
 $F^{a*}_{\mu \nu}$ and define the dual symmetry. If the corresponding $U(1)$ symmetries are Higgs
 broken magnetic charges condense in the vacuum and there is dual superconductivity (Confinement).
 If the symmetries are exact the vacuum is normal and chromoelectric charges deconfined.
   An operator $\mu$ can be constructed carrying non zero magnetic charge, and its $vev$ $\langle \mu \rangle$ can be used as an order parameter for confinement \cite{dpf}\cite{dlmp1}\cite{dlmp2}, i.e. as a detector of monopole condensation.
   \section{The phase diagram.}
   \label{3}
    A transition is a rapid change in physics at some value $T_c$ of some parameter say the temperature  $T$.  A transition shows up as a peak in susceptibilities, which are the derivatives of observables  with respect to $T$.  For example a peak in the specific heath $C_V$ is a rapid change in the heath content.
    A transition is called a crossover if no discontinuity develops at $T_c$ as the volume $V$ goes to infinity, it is named first order if some first derivative of the free energy diverges , e.g. if the free energy
    itself has a discontinuity at $T_c$ and $C_V$ diverges as $V\to \infty$.
    
    Stating that a transition is a crossover is equivalent to verify that the free energy is analytic trough $T_c$ , and this cannot be done on the basis of any numerical calculations with a finite volume and a finite resolution. It can sometime be done with the help of some theory. A classical example is the chiral transition at small quark masses in $N_f=2$ $QCD$\cite{pw}. Assuming that the relevant degrees of freedom at the chiral-deconfinement transition are the chiral ones, on the basis of renormalization group arguments one can say that either the chiral transition is second order in the universality class of $O(4)$, and then the transition is a cross-over at small non zero masses, or it is first order, and then it stays first order at small masses. In the first case a tricritical point is predicted at finite density, whose existence can be checked experimentally in heavy ion collisions\cite{ss} ; no tricritical  point exists in the second case.  Finite size scaling analysis has been performed by many groups\cite{fss}, but none finds evidence for second order $O(4)$.
    
    If the correlation lengths are large compared to lattice spacing scale invariance holds and one expects  for the volume dependence e.g. of the specific heath the following scaling law\cite{ddp}\cite{cddp}
    \begin{equation}
    C_V - C_0 \approx L_s^{\alpha\over \nu}\Phi_C( \tau L_s^{1\over \nu}, m L_s^{y_h})
    \end{equation}
    
    Here $L_s$ is the spatial size of the lattice, $\tau= (1-{T \over T_c})$ the reduced temperature and $\alpha$, $\nu$ and $y_h$ are critical indexes which are specific of the order and universality class of the transition. For second order $O(4)$  $\alpha=-.24$, ${1\over \nu} =1.34$, $y_h=1.48$.
    For weak first order $\alpha=1$, ${1\over \nu} =3$, $y_h=3$.
    Eq.(1) can be tested on lattice data either by keeping the second variable of the function $\Phi_C$ fixed, by choosing $m$ and $L_s$ such that  $m L_s^{y_h}$ has a fixed value, say $K$ for a given assumption ($y_h$) on the universality class; or by keeping the first variable fixed and checking the dependence on the second one\cite{ddp}. One has in the first case
    \begin{equation}
    (C_V-C_0)/L_s^{\alpha\over \nu} \approx  \Phi_C(\tau L_s^{1\over \nu},K)
    \end{equation}
    in the second case, at large values of $m L_s^{y_h}$\cite{cddp}, one has for second order O(4)
    \begin{equation}
    (C_V-C_0)\approx m^{.13}f_C(\tau L_s^{1.35})
    \end{equation}
     Instead for weak first order 
     \begin{equation}
      (C_V-C_0)\approx  L_s^3 f^0_C(\tau L_s^3) + {1\over m} f_C^1(\tau L_s^3)
      \end{equation}
     Data on lattices $L_t=4$, $L_s=  16, 20, 24, 32$ do not agree with the scaling Eq.(2) with the choice $O(4)$, they do with the choice weak 1st order \cite{ddp}\cite{cddp}.  Also Eq(3) is not satisfied by $O(4)$. Eq.(4) instead is obeyed, but the first term, which is typical of first order looks to be negligible at present volumes, implying that the transition is too weak to observe a growth proportional to the volume at presently available volumes.  Moreover, with $L_t=4$ the lattice at the phase transition is rather coarse, and a check should be done with smaller lattice spacings. Evidence for the existence of the first term of Eq.(4)
     is needed to make a definite statement on first order.  No definite evidence exists by now for a cross-over.    
    
 %which are conserved due to the antisymmetry of %the dual tensor $F^{a\star}_{\mu \nu}$ \cite{dfp}.   
% \appendix
%
% \section{}
% \label{}
%

%
\end{document}